\documentclass[MSNbibl,nameyear,dvips]{arxstspdf}
\usepackage{flushend}
\usepackage{stfloats}
\usepackage{graphicx}

\volume{27}
\issue{3}
\pubyear{2012}
\firstpage{434}
\lastpage{445}
\doi{10.1214/12-STS387} %kopijuoti is PTS
\makeatletter
\newtheorem{them}{Theorem}
\makeatother

\begin{document}
\begin{frontmatter}

\title{Pao-Lu Hsu (Xu, Bao-lu): The~Grandparent of Probability and
Statistics in China}%\thanksref{T1}
% kai straipsnis turi susijusiu diskusiju ir rejoinder'iu
%rejoinder at \relateddoi{r}{10.1214/00-STSXXXX}.}
\runtitle{Pao-Lu Hsu}

\begin{aug}
\author[a]{\fnms{Dayue} \snm{Chen}\ead[label=e1]{dayue@pku.edu.cn}}
\and
\author[b]{\fnms{Ingram} \snm{Olkin}\corref{}\ead[label=e2]{olkin@stanford.edu}}
\runauthor{D. Chen and I. Olkin}

\affiliation{Peking University and Stanford University}

\address[a]{Dayue Chen is Professor of Mathematics,
School of Mathematical Sciences, Peking University,
Beijing, China 100871 \printead{e1}.}
\address[b]{Ingram Olkin is Professor Emeritus of Statistics and Education,
Department of Statistics, Stanford
University, Stanford, California 94305, USA \printead{e2}.}

\end{aug}

% ABSTRACT
%
\begin{abstract}
The years 1910--1911 are auspicious years in Chinese mathematics with
the births of Pao-Lu Hsu, Luo-Keng Hua and Shiing-Shen Chern. These
three began the development of modern mathematics in China: Hsu in
probability and statistics, Hua in number theory, and Chern in
differential geometry. We here review some facts about the life
of P.-L. Hsu which have been uncovered recently, and then discuss
some of his contributions. We have drawn heavily on three papers in
the 1979 \textit{Annals of Statistics} (volume 7, pages 467--483)
by T. W. Anderson, K. L. Chung and E. L. Lehmann, as well as an article
by Jiang Ze-Han and Duan Xue-Fu in Hsu's collected papers.
\end{abstract}

% KEYWORDS
%
\begin{keyword}
\kwd{Multivariate analysis}
\kwd{Wishart distribution}
\kwd{Student's $t$}
\kwd{Hotelling's $T^2$}
\kwd{determinantal equation}
\kwd{eigenvalues}
\kwd{design of experiments}
\kwd{mathematics in China}.
\end{keyword}

\end{frontmatter}

%s1 ###
\section{Hsu's life}\vspace*{3pt}

Pao-Lu Hsu was born in Beijing on September~1, 1910, ``into a Mandarin
family from the famed lake city of Hangchow'' in the Zhejiang Province
in Eastern China. His family was well educated: not only his father,
but also his grandfather, his great-grandfather and the father of his
great-grandfather, as well as their brothers and brothers-in-law. This
reflected the tradition in old China of excelling in local exams,
provincial exams and finally in national exams. The tradition was
abandoned after China lost the war with Japan in 1895 and turned to
Western methods.

Hsu was the youngest of seven children, with two brothers and four
sisters. His father died when he was~14. During his childhood, Hsu
moved from Beijing to Tianjin to Hangzhou and back to Beijing. His
early education was with private tutors, a luxury that few people could
afford at that time. He began attending school at the age of 15, and
enrolled at Yenching University when he was 18. He first studied
chemistry, then decided to study mathematics and transferred to
Tsinghua University in 1929. Both Yenching and Tsinghua universities
had connections with universities in the United States. For example,
the Yenching--Harvard Institute, founded in 1928, was designed to foster
education in the humanities and social sciences in Asia.

%f1 ###
\begin{figure*}
\centering

\includegraphics{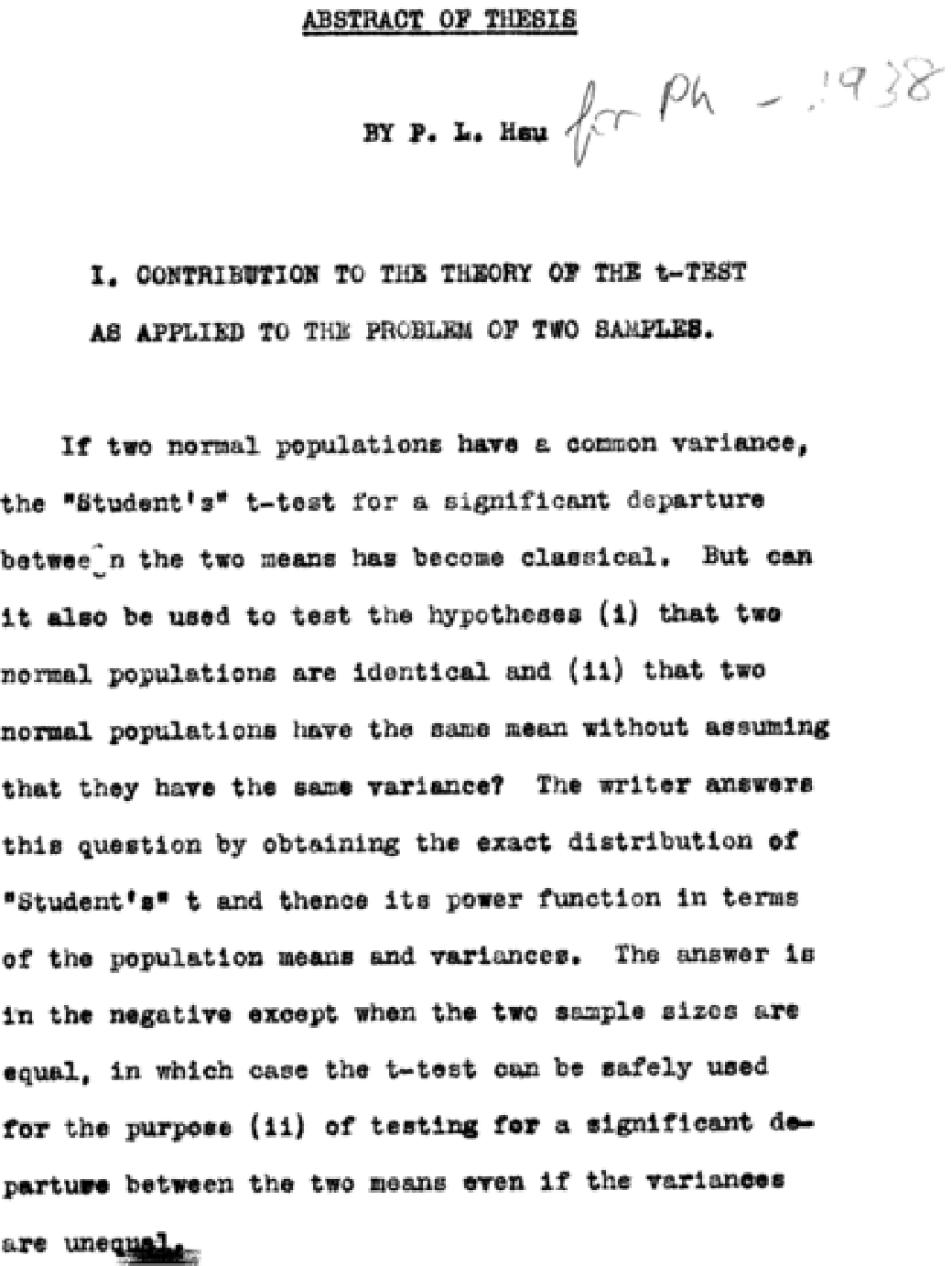}

\caption{Abstract of Pao-Lu Hsu's 1938 thesis, part I: ``Contribution
to the theory of the $t$-test as applied to the problem of two samples.''}\label{fig1}
\end{figure*}

%f2 ###
\begin{figure}
\centering

\includegraphics{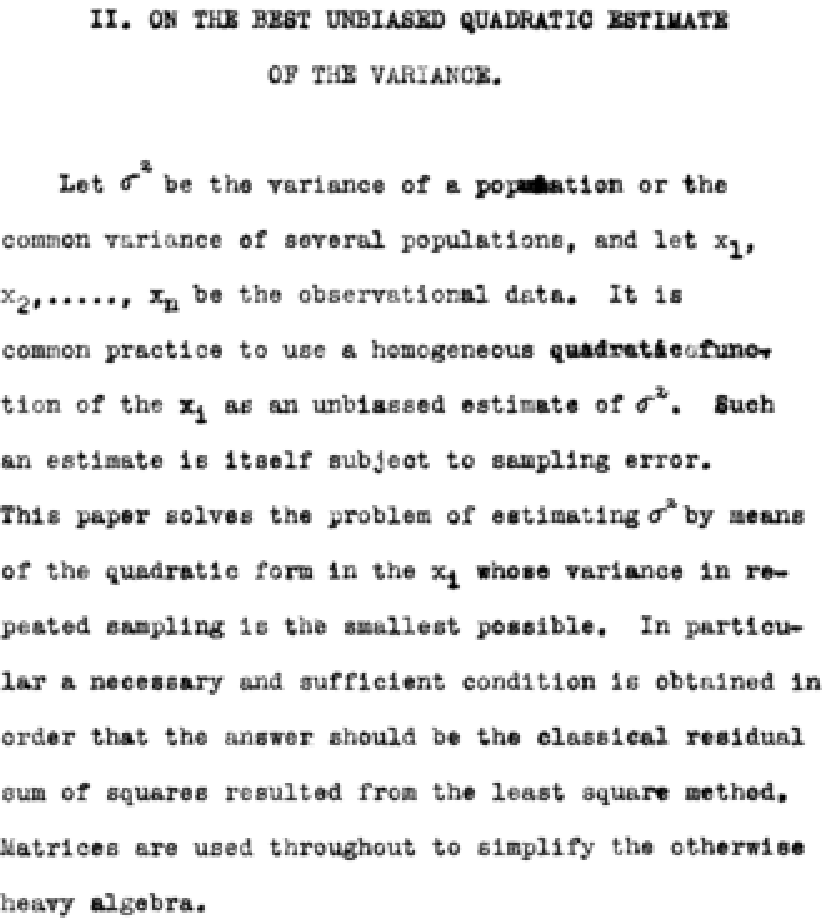}

\caption{Abstract of Pao-Lu Hsu's 1938 thesis, part II: ``On the best
unbiased quadratic estimate of the variance.''}\label{fig2}
\end{figure}

Hsu received a bachelor of science degree from Tsinghua University, and
then went to Peking University, where he was an assistant in the
Department of Mathematics. He passed the examination in 1936, after
which he went to the University of London to continue his studies. He
served as a lecturer, obtaining a Ph.D. in 1938 and a Sc.D. in 1940.
Thus, he remained in England for the four years 1936--1940. However, it
is known that he spent some time in Paris (perhaps during the academic
year 1939--1940) to study with Jacques Hadamard. Hsu then returned to
China, which was again at war with Japan. (The Second Sino--Japanese War
lasted from 1937 to 1945.) After his return, Hsu was appointed as
professor at Peking University, which was relocated to Kunming during
World War II.

Constance Reid writes in her biography of Jerzy Neyman: ``The most
outstanding of Neyman's students at that time (eds. 1937--1938) was a
Chinese, \ldots P. L. Hsu. (Neyman expresses to me his admiration for
Hsu with a Polish phrase which he trans\-lates---with a little bow and
a gracious wave of the hand---as \textit{Please} sit down!)'' (\cite
{reid}, page 153).
The Neyman biography further notes that, ``Hsu was,
in Neyman's opinion, absolutely on a level with Wald---they were the
two outstanding statisticians in the generation coming up!'' Hsu was
invited to lecture at Berkeley for six months with an appointment for
the following year. Harold Hotelling was at Columbia at this time and
suggested that Columbia and Berkeley join together to bring Hsu to each
university for a semester. (Hsu was also sought after by Chicago and
Yale.) Hsu accepted the joint offer and indicated that he preferred to
first visit the West Coast.

Erich Lehmann in his autobiography lists Hsu as one of his three Ph.D.
godfathers. At that time (1945) Lehmann was a doctoral student at
Berkeley, and Neyman asked Hsu to give him a thesis topic:

\begin{quote}
Within a few days, Hsu presented me with a new possible topic: applying
methods of Neyman, Scheff\'e and himself to some situations for which
they had not been tried before. Hsu then got me started on this line of
work. In a letter of January 24 to Neyman, about which I learned only
much later, he wrote: ``I have passed the problem of testing for
independence between successive observations to Erich for his doctoral
thesis. Will do all I had done independently,and then add a new part
which I have not done. I hope this scheme will meet with your approval,
so that Erich can look forward to the degree with certainty.''

\quad This was an act of greatest generosity. Hsu made me a present of work
he had planned to do himself and on which he had already obtained some
results. I had hoped to see him on his return to Berkeley after the
term at Columbia. However, this was not to be; in fact I never saw him
again. (\cite{lehmann}, page 39)
\end{quote}

When Hotelling moved to the University of North Carolina in 1946 to
found a department of mathematical statistics, he offered Hsu an
associate professorship. Hsu accepted the offer and spent the period
1946--1947 at Chapel Hill, but the pull to return to China was too
strong, and he returned in the summer of 1947.\vadjust{\goodbreak} He was committed to
China, and wanted to participate in ``the emerging new society in his
homeland.'' On a trip across country (1946--1947) Neyman visited
Hotelling in Chapel Hill and saw Hsu again, whom he hoped to entice to
Berkeley. ``He found the Chinese scholar miserably unhappy,
disappointed in love, and desiring only to return to his native land.''
(\cite{reid}, page 214)

%f3 ###
\begin{figure*}
\centering

\includegraphics{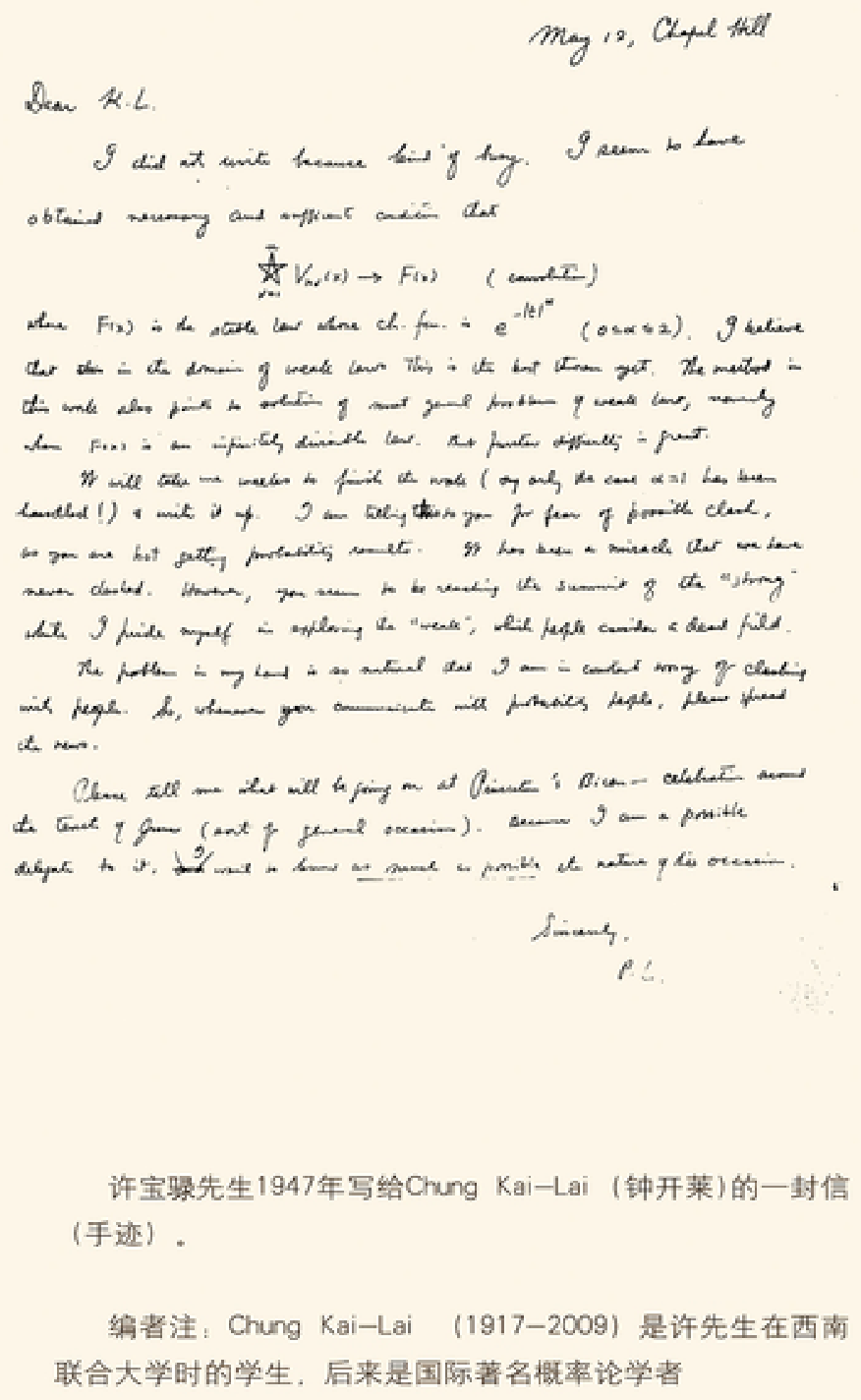}

\caption{Pao-Lu Hsu's handwritten letter to Kai-Lai Chung in 1947.
K.-L. Chung (1917--2009) was Professor Hsu's student during the Second
World War and later became a well-known probabilist.}\label{photo1}
\end{figure*}

After his return to China, Hsu's research was unknown in the West.
``Apart from his published papers and a few remarks given us by an old
friend, we are unable to obtain further information about Hsu's life
and work in the twenty-some years he lived in Peking'' (\cite{1}). His
colleagues at Peking University did not see him easily either. As
reported by Boju Jiang, who joined the department as a faculty member
in 1957 (but did not meet Hsu until 1968), ``Mr. Hsu was essentially a
legendary hero, somewhat mysterious to us.''

Hsu was the first teacher to offer courses in probability and
statistics in China, from the early 1940s in Kunming. Kai-Lai Chung was
a teaching assistant at that time, and became interested in probability
by taking courses and discussing research with Hsu. For this reason
Chung always regarded himself as a student of Hsu. Other students
included Shou-Jen Wang, L.C. Hsu and Chin-long Chiang. Under Hsu's
supervision, Zhong-Zhe Zhao completed graduate study in 1951 and was
the first graduate student to major in probability in China. After
three lectures given in the fall of 1955, Hsu could no longer teach in
a classroom due to his poor health.

In 1956, probability and statistics (together with computational
mathematics and differential equations) were identified as key subjects
of mathematics to be developed with high priority in China. Only a~few
Chinese researchers knew probability and statistics at~that time, and
in order to produce qualified teachers at an accelerated pace, a~special program was created at Peking University, with 34 juniors from
PKU, 10 juniors from Nankai University in Tianjin and 10 juniors from
Sun Yat-sen University in Guangzhou (Canton). In addition, some 10
teachers came from all over the country to audit the courses.
Instructors were brought from the Chinese Academy of Science and Sun
Yat-sen University. Hsu was a great teacher, and served as leader of
the program. The curriculum he created there later became the national
standard. Textbooks were compiled quickly, some based on the notes from
his lectures and some translated from Russian. After a~two-year
training period, students were dispatched to other universities\vadjust{\goodbreak} to
teach probability and statistics. In some sense, all Chinese
probabilists and statisticians are students or grand-students of Hsu.
On teaching, he once said: ``One can feel proud to be the advisor of a
Nobel laureate. It means nothing just to be a student of a Nobel laureate.''

%f4 ###
\begin{figure*}
\centering

\includegraphics{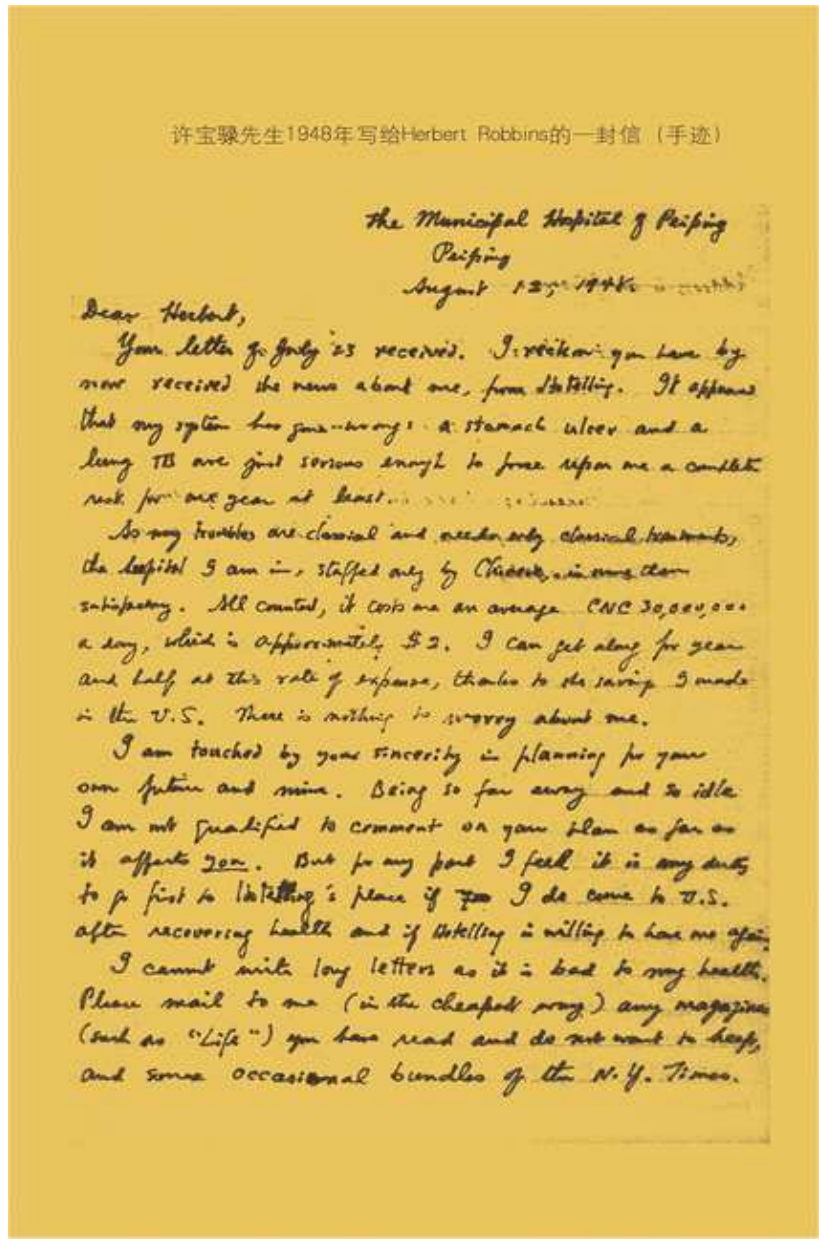}

\caption{Pao-Lu Hsu's handwritten letter to Herbert Robbins in 1948
(page 1 of 2).}\label{photo2}\vspace*{-3pt}
\end{figure*}

\begin{figure*}
\centering

\includegraphics{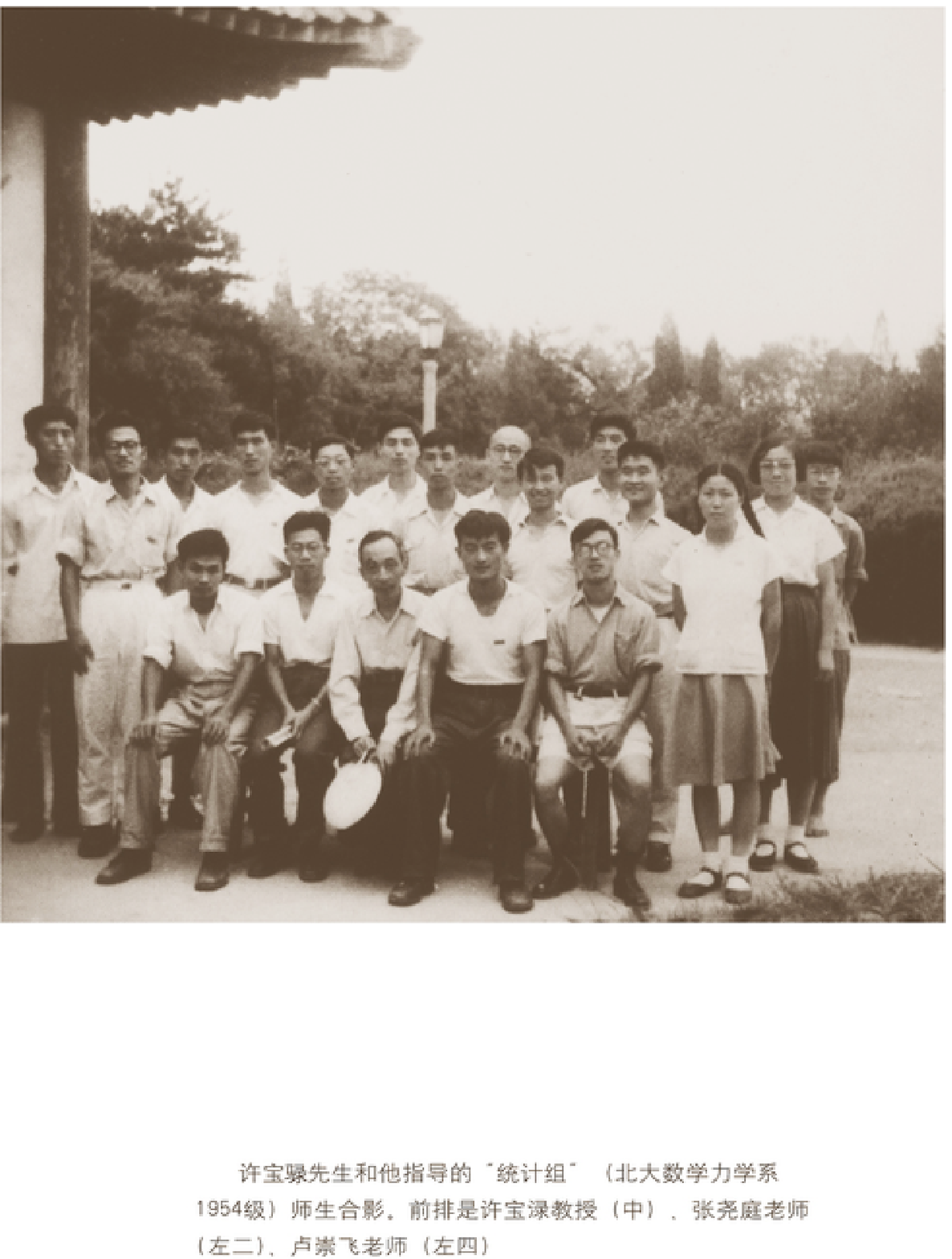}

\caption{Pao-Lu Hsu and his group of statistics students in the
Department of Mathematics and Mechanics at Peking University in 1954.
In the front row are Professor Hsu (center) and his assistants Yao-Ting
Zhang (second from left) and Chong-Fei Lu (second from right).}\label{photo4}
\end{figure*}

\begin{figure*}
\centering

\includegraphics{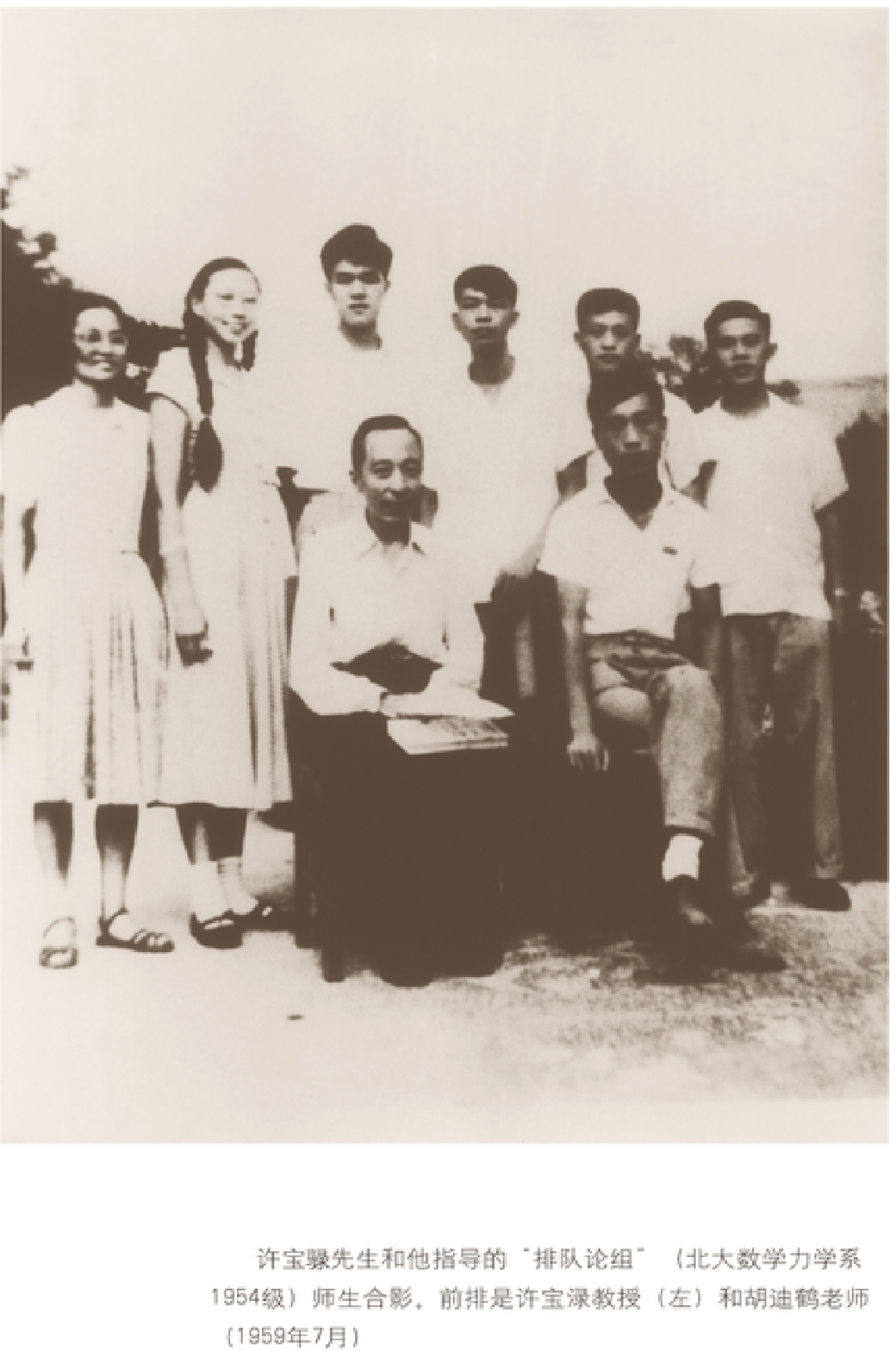}

\caption{Professor Pao-Lu Hsu (left), Professor Deehe Hu (right) and
the Queueing Theory Discussion Group (1954 Entering Class, Department
of Mathematics and Mechanics, Peking University), circa July,
1959.}\label{photo5}\vspace*{-3pt}
\end{figure*}

Hsu continued his teaching by running seminars at his home. He began
the practice informally in the early 1950s, and continued on a regular
basis for eight consecutive years until 1964. This was very much in the
Chinese tradition of private education in which students were required
to learn by themselves and to present their findings each week for
evaluation by the advisor. Quite often the seminar became a small class
taught by Hsu himself. Only a few people were fortunate to serve as his
apprentices. During the peak period, he ran three seminars
a week in the living room of his one-bedroom apartment on
campus, which was about 140 square feet in size. Seminar participants
were selected by Hsu himself; in contrast, his graduate students were
assigned to him. Among the few photographs available today are two
taken with students of his seminars in 1959 (Figures~\ref{photo4} and~\ref
{photo5}). The topics of the seminars covered a wide range:
mathematical statistics, limit theory, Markov processes, stationary
processes, experimental designs, sampling techniques, order statistics,
and topology. Research conducted by members of the seminars represented
the first coordinated efforts in probability and statistics in China.
Hsu created pen names such as Ban-cheng, Ban-guo and Ban-ji for the
students to write and submit joint papers. (The Chinese word ``Ban''
means class.) From 1958 to 1962 he also supervised six graduate
students, including Yongquan Yin.

%f5 ###
\begin{figure*}
\centering

\includegraphics{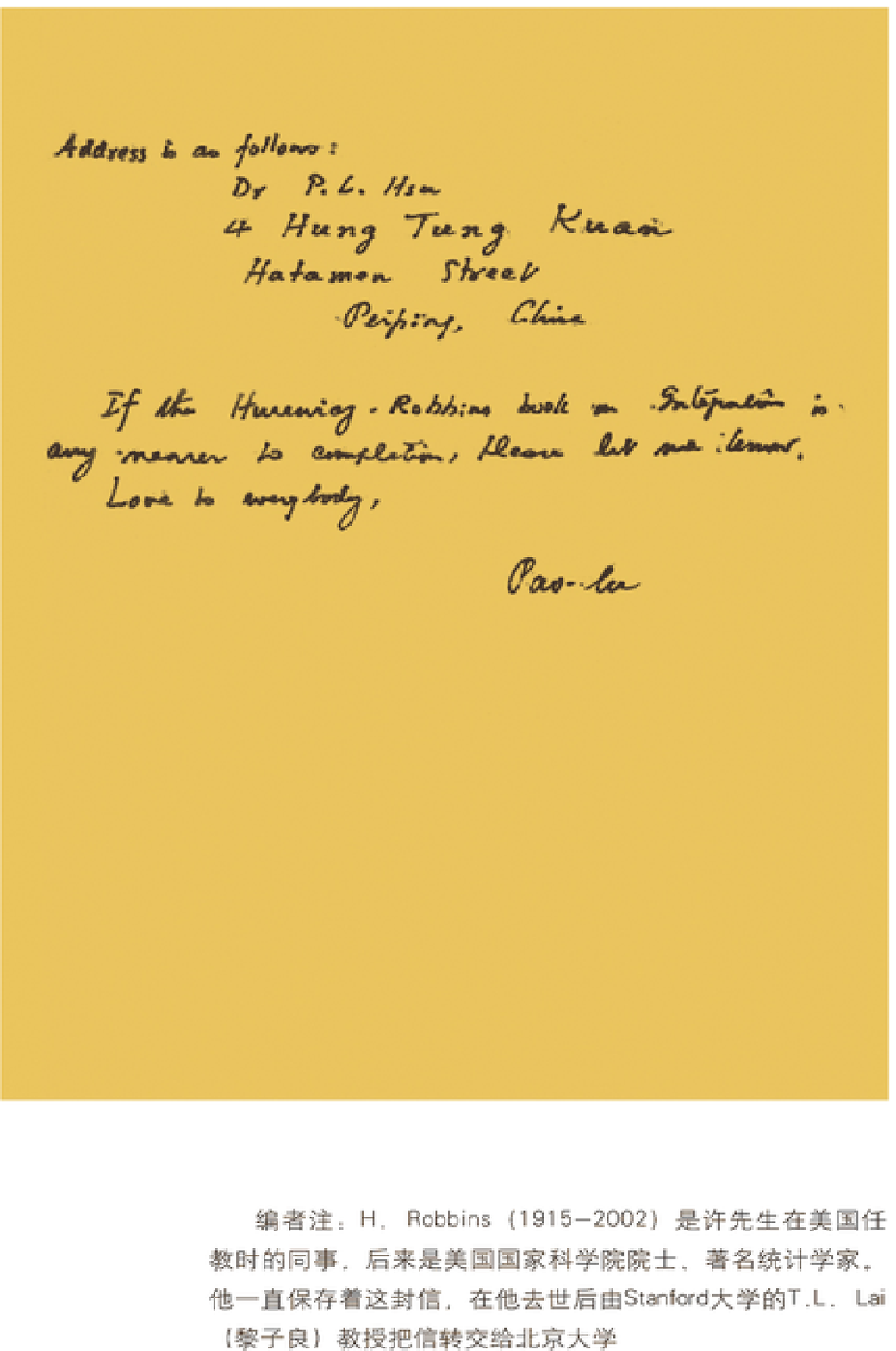}

\caption{Hsu's 1948 letter to Robbins (page 2 of 2). Herbert Robbins
(1915--2002) was Professor Hsu's colleague at Columbia University. He
was an eminent statistician and a member of the US National Academy of
Sciences. Robbins had kept this letter for a long time, and after his
death the letter was sent to Peking University by Professor T. L. Lai of
Stanford University.}\label{photo3}
\end{figure*}

The reader is reminded that China and the Soviet Union were in a
``honeymoon'' period in the 1950s, and China learned much from the
Soviet Union. Following the Soviet pattern, a subdivision within the
mathematics department was set up at Peking University in 1956,\vadjust{\goodbreak} named
\textit{Teaching and Research Unit of Probability Theory and
Mathematical Statistics}. This was the very first of its kind in China,
and it evolved into an independent department in 1985. As the founding
director of the unit, Hsu watched over the career development of young
colleagues, because most members only had undergraduate training. He
also organized scientific exchanges with foreign colleagues, for
example, Marek Fisz and Kazimierz Urbanik from Poland in 1957 and
Eugene Dynkin from the USSR in 1958. Careful preparation was made
before each visit in order to better understand the forthcoming
lectures. Several months before these visits, Hsu would assign related
papers for young faculty members and students to study.

However, although Hsu himself was immune from politics, Hsu's efforts
were discounted. Seminars were ended unexpectedly because students were
required to devote themselves fully to a political movement. Graduate
students were selected based on political criteria and were assigned to
professors, without much consideration of the academic interest and
ability of the student. Some were not well prepared for graduate study.
Hsu must have been annoyed by the requirement to submit a research plan
each year, simply because everything had to be part of a planned
economy. His solution was to propose his new papers as the research
plan for the next year.

Shortly before the Communist victory in 1949, Hsu and most other
professors declined the offer of Chiang Kai-shek to airlift them to
southern China. He even sent a telegram to a foreign friend saying
``\ldots am happy after liberation.'' However, when Hsu returned from
the US in 1947, China was in the middle of their civil war. His return
from the UK to China in 1940 had been even worse. At that time, China
was involved in WWII and living conditions in Kunming were miserable.
On May 4th of 1919, students of Peking University demonstrated in
Tian-an-men Square, crying out for the adoption of principles of
democracy and science. (This was the first of a number of Tian-an-men
Square protests, the most recent occurring in 1989.) The May 4th
Movement was a turning point in the modern history of China. It is
reasonable to assume that Hsu, like many Chinese intellectuals of his
generation, wished to build a strong country by introducing science to
China. This may also explain in part why Hsu submitted papers to
Chinese journals.

In the 1950s Hsu wanted to create a Chinese journal in probability and
statistics, as a launch pad for young researchers to publish their
papers. He was prepared to subsidize the journal even with his own
money. Hsu had a tremendous linguistic talent, with a command of
English, German, French and Russian. He learned Russian by himself, and
helped correct several textbooks by Alexander Khintchine and Vyacheslav
Stepanov, translated from their original editions. Indeed, the last
accomplishment of his life was to proofread a set of manuscripts
scheduled for completion in one month. He looked at the task and said
he could do it in ten days; he finished the job in a little over nine days.

Hsu's health had been fragile since he was young. He was 5 feet, 9
inches tall, but only weighed 88 pounds at his maximum. Because of his
light weight, he was disqualified for a government fellowship to study
abroad in 1933. According to medical records, he was hospitalized in
1948 and in the early 1950s, and recuperated in hospital from illnesses
in 1933 and 1957. In a letter to Herbert Robbins in 1948, Hsu wrote,
``It appears that my system has gone wrong, a stomach ulcer and a lung
TB are just serious enough to force upon me a complete rest for one
year at least.'' In his final decade or so he was essentially confined
to bed, where he continued to read and write. He never married and
lived alone.

For his important contributions, Pao-Lu Hsu was one of five
mathematicians elected as Academicians in 1948. He was elected as an
Academician again in 1955, along with eight other mathematicians. He
was a Fellow of the Institute of Mathematical Statistics. After his
death, memorial meetings were held every ten years at Peking
University. In 2010, a~me\-morial collection\vadjust{\goodbreak} of papers was published,
a~bronze statue was dedicated, an international conference on probability
and statistics was held in July, and~an official commemoration was held
on his centennial birthday. The Pao-Lu Hsu Lecture Series was launch\-ed
at Peking University in 2009 with a roster of distinguished speakers.
Tsinghua University recently also inaugurated the Pao-Lu Hsu
Distinguished Lecture in Statistics and Probability, with Brad Efron as
the first speaker. The P.-L. Hsu Conference on Statistical Machine
Learning is now in its third year. The International Chinese
Statistical Society (ICSA) has announced its intention to set up the
P.-L. Hsu Award. A memorial webpage will be built as part of PKU's
Department of Probability and Statistics web site,
\url{http://www.stat.pku.edu.cn}.

%s2 ###
\section{Hsu's Research}\label{sec2}

Some insight into Hsu's views about research can be gleaned from
comments made to his students, such as the following:
\begin{itemize}
\item``The merit of a paper is not just to get published, but is
realized when it is cited repeatedly by others.''
\item``A good author should show the simplicity.''
\item``I do not want to become famous because my paper appears in a
well-established journal. I wish a~journal to be well established
because my paper appears in that journal.''
\end{itemize}

Pao-Lu Hsu authored 41 papers and three books, on a wide range of
topics: limit theorems, random matrices, Markov chains, experimental
designs, characteristic functions. Almost all his papers were singly
authored. He published with only three coauthors: Kai Lai Chung,
Tsai-han Kiang and Herbert Robbins. Hsu used the pen name Ban-cheng to
publish papers with his students. As demonstrated in Table~\ref{table1}, his
peak performance was from 1938 to 1947. In a ten-year period he wrote
22 papers, all in English. For the next 15 years (1949--1964), he wrote
13 papers of which 11 articles were published in Chinese journals. Some
of his manuscripts were published posthumously, with the assistance of
his students.

%t1 ###
\begin{table}
\tablewidth=240pt
\tabcolsep=0pt
  \caption{Hsu's publications}\label{table1}
\begin{tabular*}{240pt}{@{\extracolsep{\fill}}llccl@{}}
\hline
& & \multicolumn{2}{c}{\textbf{Journals}} &
\\
\ccline{3-4}
\textbf{Period}& \textbf{Residence}& \textbf{Chinese}& \textbf{Int'l}& \textbf{Topics}
\\
\hline
1935& Beijing& \phantom{0}1& 1& topology
\\
1938--1940& London& \phantom{0}1& 7& statistics
\\
1941--1945& Kunming& \phantom{0}1& 9& random matrices, \\
&&&& limiting distributions
\\
1946--1947& USA& \phantom{0}0& 4& matrices, \\
&&&& complete convergence
\\
1949--1964& Beijing& 11& 2& characteristic
functions,\\
 &&&& experimental design, \\
 &&&&matrix transformations, \\
 &&&&Markov
processes,\\
&&&& limiting distributions
\\
1968& Beijing& \phantom{0}0& 1& limiting distributions
\\
Posthumously& & \phantom{0}3& 0& Markov chain, \\
&&&&random matrices, coding
\\
\hline
\end{tabular*}
\end{table}

%s3 ###
\section{Hsu's Research in Multivariate Analysis}\label{sec3}

In addition to probability and statistics, Hsu lectured on topology,
matrix theory and analysis. He did not have access to outside
literature, and provided new proofs of some known results. He also
obtained new results which still remain hidden.

%s3.1 ###
\subsection{The Wishart Distribution}\label{sec31}

The Wishart distribution was a focal point in Hsu's early work. He
generated new derivations and obtained the distribution of the
eigenvalues of the sample covariance matix and the canonical correlations.

The joint distribution of the elements of the covariance matrix
obtained from a sample of $p$-variate standard normal variates was
obtained by Fisher in 1915 for the special case $p=2$, and by Wishart
in 1928 for general $p$. (Although stated for standard normal
variables, by a simple transformation the variates can have a
covariance matrix $\Sigma$.) What is tantalizing about this
distribution is that the starting point is a set of $pn$ ($p\leq n$)
random variables in the $p\times n$ sample matrix $X$, and the ending
point is a set of $p(p+1)/2$ random variables in the matrix $S=XX'$.
There are a number of routes and methods that might be used to make
this transition. Wishart's original derivation was a geometric argument
and later in 1933, together with Bartlett, he gave a derivation using
characteristic functions. In 1937, Mahalanobis, Bose and Roy gave a
geometric derivation of rectangular coordinates (described below),
which is a stepping stone to deriving the Wishart distribution.

In 1939, Hsu gave a new derivation of the Wishart density using a
clever inductive argument. The case of $p=1$ reduces to the chi-square
distribution. The essence of the induction is to go from $p-1$ to $p$
variables. Here Hsu uses a multivariate tranformation, an area that he
later developed.

In 1940 Hsu gave an algebraic derivation of rectangular
coordinates\vadjust{\goodbreak}
which led to a general result. Suppose the joint density $p(X)$ of the
$qm$ elements of the random $q\times m$ matrix $X$ is of the functional
form $p(X)=f(XX')=f(S)$, where $S=XX'$. Let $S=TT'$, where $T$ is lower
triangular. The elements of $T$ are called \textit{rectangular
coordinates}. Hsu gives a~detailed proof that such a factorization
exists; this is an example in which Hsu gives a new proof of a~known
theorem, in this case by Toeplitz in 1907. Hsu next obtains the
Jacobian of the transformation, which leads to the joint density of the
elements of $T$:
%
%e1 ###
\begin{eqnarray}
c(q,m)\prod_1^pt_{ii}^{m-i}f(TT'),\nonumber
\\
\eqntext{0<t_{ii}, -\infty<t_{ij}<\infty,i\neq j,}
\end{eqnarray}
where $c(q,m)$ is a normalizing constant explicitly determined.

%s3.2 ###
\subsection{Roots of a Determinantal Equation}\label{sec32}

In two papers of 1933 and 1936, Hotelling opened the door to
consideration of the distribution of the roots of a determinantal
equation $|A-\theta I|=0$ or $|A-\theta(A+B)|=0$, where
$p$-dimensional random matrices
$A$ and $B$ are
independently distributed, each having a standard Wishart distribution.

%f6 ###

In the first case suppose that the random positive definite matrix $A$
has a density $f(A)$ that is orthogonally invariant, that is,
$f(A)=g(\theta_1,\ldots,\theta_p)$, where the $\theta$'s are the
eigenvalues of $A$. Make the transformation $A=\Delta D_\theta\Delta'$,
where $\Delta$ is orthogonal and $D_\theta=\mathrm{diag}(\theta_1,\ldots
,\theta_p)$ to yield the joint density of $\Delta$ and $\theta$.
Integration over the orthogonal group yields the joint distribution of
the eigenvalues. At that time (1953), this integration was not very
well known. Here again, Hsu found a clever way to carry out this
integration. He showed that every orthogonal matrix $\Gamma$ has a
representation in terms of a skew-symmetric matrix $Y$, namely, $\Gamma
=2(I+Y)^{-1}-I$. The beauty of this transformation is that whereas the
$p(p-1)/2$ variables in $\Gamma$ are not explicit, they are explicit in
$Y$. Hsu then shows how to carry out the integration to yield a general result.
\begin{them}
If $S$ is a random $p$-dimensional positive definite matrix with an
orthogonally invariant density $f(S)=g(\theta_1,\ldots,\theta_p)$,
where $\theta_1\geq\cdots\geq\theta_p>0$ are the eigenvalues of $S$,
then the joint distribution of the eigenvalues is
\[
c(p)\prod_{i<j}(\theta_i-\theta_j) g(\theta_1,\ldots,\theta_p),
\]
where $c(p)=\pi^{p(p+1)/4}/\prod_1^p\Gamma(i/2)$.
\end{them}

To obtain the distribution of the roots of $|A-\theta(A+B)|=0$, Hsu
transforms $(A,B)$ to $(W,\varphi)$ by $A=WD_\varphi W', B=WW'$, where
$D_\varphi=\mathrm{diag}(\varphi_1,\ldots,\break\varphi_p)$. Note that
$\varphi_i=\theta_i/(1-\theta_i)$. The key problem is to evaluate the
Jacobian of the transformation. Here Hsu gave an explicit derivation
for $p=3$ but stated the general result, later given in the exposition
by Deemer and Olkin (\citeyear{deemer}). Hsu obtained the result for $p<n_1,n_2$ and
$n_1<p<n_2$, where $n_1$ and $n_2$ are the sample sizes that lead to
$A$ and $B$.

%f7 ###

%f8 ###
\begin{figure*}
\centering

\includegraphics{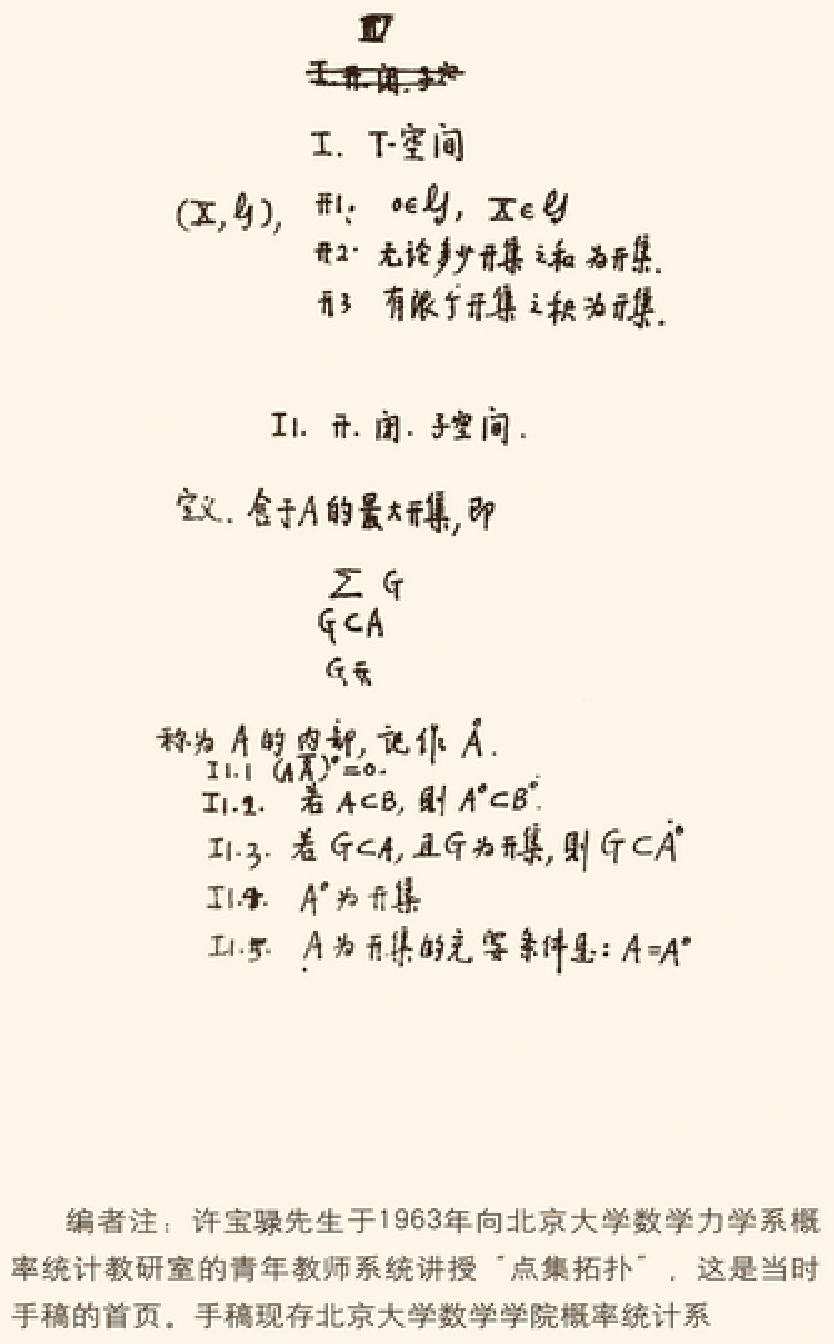}

\caption{First page of handwritten notes, kept at the Department of
Probability and Statistics of the School of Mathematical Sciences, of
Professor Pao-Lu Hsu's lectures on point-set topology in 1963 at the
Department of Mathematics and Mechanics, Peking
University.}\label{photo6}\vspace*{-3pt}
\end{figure*}

In the description above, both $A$ and $B$ have a~central Wishart
distribution. Hsu (\citeyear{hsu}) tackles the case that $B$ has a central
Wishart distribution but~$A$ has a noncentral distribution. This
distribution is complicated and Hsu obtains asymptotic results. See
Anderson (\citeyear{3}) for a more detailed description.

%s3.3 ###
\subsection{Student's $t$ and Hotelling's $T^2$ Distributions}\label{newsec33}

Hsu's first statistical paper was in 1938, and in it he obtained the
distribution of the square of the Student's $t$-statistic in which the
denominator is a~linear combination, $as_1^2+bs_2^2$, of the variances
in the two underlying samples. This is now called the Behrens--Fisher problem.

Hotelling obtained the null distribution of the multivariate version of
the Student's $t$-statistic in 1931, and in 1938 Hsu obtained its
noncentral distribution. This may be the earliest noncentral
distribution in multivariate analysis, and was very much in the spirit
of the Neyman--Pearson\vadjust{\goodbreak} view. (Recall that Neyman invited Hsu to the
first Berkeley Symposium in 1945. His paper was titled ``The limiting
distribution of functions of sample means and application to testing
hypotheses.'')

Hsu also wrote about power functions of several multivariate tests, and
provided a canonical form for the multivariate analysis of variance.
Here the mean of the random $p\times m$ matrix $Y$ is $EY=\Theta$, the
mean of the random $p\times n$ matrix $Z$ is $EZ=0$; the rows of $Y$
and $Z$ are multivariate normal with a common covariance matrix $\Sigma
$. The hypothesis in question is $H:\Theta=0$.

%s3.4 ###
\subsection{Design of Experiments}\label{sec33}

Hsu entered into another realm, that of combinatorial analysis. Here he
defines a balanced incomplete block (BIB) design in terms of a (0, 1)
matrix~$T$ of dimension $v\times b$ with the properties
\[
e'T=ke', \quad Te=re \quad\mbox{and}\quad TT'=rI+(\lambda+r)J,
\]
where $e$ is the vector of ones, and $J$ is the matrix with all
elements equal to 1. Such a BIB design is denoted by the five constants
$(v,b,r,k,\lambda)$. The following is an example of his results.
Suppose that a~$\operatorname{BIB}(v,\ast,r,k,\lambda)$ exists, Hsu shows how to
obtain a~$\operatorname{BIB}(v+1,\ast,r,k,\lambda)$.

He then defines a \textit{code}
as a new $n\times w$ matrix $M$ with elements $(1,-1)$. The transmitter
sends only rows of $M$ (without repetition). Let $MM'=Q=(q_{ij})$. Hsu
obtains a number of results concerning the matrix $Q$, such as the inequality
\begin{eqnarray*}
\max_{i\neq j}q_{ij}&\geq&\frac1{n(n-1)}\sum_{i\neq j}q_{ij}
\\
&\geq&
\cases{
-w/(n-1),& if $n$ is even,
\cr
-w/n, &  if $n$ is odd.}
\end{eqnarray*}

Hsu defines a \textit{simple code} to be a matrix $M$ for which
equality is achieved, and then obtains necessary and sufficient
conditions for $M$ to be a simple code. An \textit{orthogonal code} is\vadjust{\goodbreak}
one for which $MM'=wI$, where $I$ is the identity matrix. Hsu provides
an example of a $12\times20$ orthogonal code. It is to be noted that a
$4w\times4w$ orthogonal code is a Hadamard matrix, a subject of
interest for a long time. Hsu is aware of this interest (recall that he
studied with Hadamard) so raises the following question. Suppose, by
his construction, that we can generate an $n\times4t$ orthogonal code:
is it possible to extend this to a $4t\times4t$ orthogonal code? He
provides two counterexamples, one of dimension $4\times12$ and the
other $12\times20$. Hsu's interest in codes may stem from his study
with Jacques Hademard, who connected the construction of
error-correcting codes with matrices whose elements are $+1$ or $-1$
and whose rows are orthogonal.

%s4 ###
\section{Epilogue}\label{sec4}

Hsu's last paper, ``BIB matrices, simple codes and orthogonal codes,''
was published in 1970. Zhang Yao-ting provided this introduction for
its appearance in Hsu's collected papers (page 566):

\begin{quote}
This paper was Professor Pao-Lu Hsu's last article, completed in
October 1970. He died in December of the same year. The Cultural
Revolution lasting already four years at the time had not yet ended. He
had suffered tremendously in this cala\-mity, and physically he was
paralyzed. This paper was completed when he was bedridden. The only
journal he could have access to was the \textit{Annals of Mathematical
Statistics}. It was said that when he gave this paper to Mr. H. F.
Tuan, being no longer able to speak clearly, he was using his hands to
express himself.

The material discussed in this article is closely related to the
contents covered in his early 1966 seminars on combinatorial analysis.
During the later years of his life, he was devoted to using matrices to
describe and prove results in combinatorial analysis. This article is a
typical representative of this idea.
\end{quote}

\section*{Acknowledgment}\label{sec5}

We thank Professor Tom Fearn of University College, London, for
his persistence in uncovering Hsu's history as a doctoral
candidate.\vspace*{27pt}

%
% imsref loaded by audrone.aklyte, 2012-04-12 14:01:44

\end{document}